\documentclass{sig-alternate}
\usepackage[T1]{fontenc}
\usepackage[latin9]{inputenc}
\usepackage{color}
\usepackage{verbatim}
\usepackage{amsmath}
\usepackage{graphicx}
\usepackage{wrapfig}
\usepackage{listings}
\usepackage{mdwlist}
\usepackage{url}
\usepackage{multicol}
\usepackage{enumerate}
\usepackage{caption}
\usepackage{ifpdf}

\ifpdf
   \usepackage{etoolbox}
   \patchcmd{\maketitle}{\@copyrightspace}{}{}{}
   \usepackage{subcaption}
\else
   \usepackage{subfigure}
\fi

\begin{document}
\conferenceinfo{Gold Coast}{'14 Gold Coast, Australia}

\lstset{ %
language=Java,                
basicstyle=\scriptsize,       
numbers=left,                   
numberstyle=\scriptsize,      
stepnumber=1,                   
numbersep=5pt,                  
backgroundcolor=\color{white},  
showspaces=false,               
showstringspaces=false,         
showtabs=false,                 
frame=single,           
tabsize=2,          
captionpos=b,           
breaklines=true,        
breakatwhitespace=false,    
escapeinside={\%*}{*)}          
}

\linespread{2}

\title{Bullseye: Passage Retrieval and Highlighting for Scholarly Search}
\title{Bullseye: Structured Passage Retrieval and Document Highlighting for Scholarly Search}

 \numberofauthors{2}
 \author{
   \alignauthor Xi Zheng \\
   \affaddr{The Center for Advanced Research in Software Engineering}\\
   \affaddr{The Univeristy of Texas at Austin}\\
   \email{jameszhengxi@utexas.edu}
   \alignauthor Akanksha Bansal, Matt Lease\\
   \affaddr{Department of Computer Science}\\
   \affaddr{The Univeristy of Texas at Austin}\\
   \email{akankshabansal90@gmail.com,ml@utexas.edu}
 }
\maketitle

\begin{abstract}
We present the {\em Bullseye} system for scholarly search. Given a collection of research papers, Bullseye: 1) identifies relevant passages using any off-the-shelf algorithm; 2) automatically detects document structure and restricts retrieved passages to user-specified sections; and 3) highlights those passages for each PDF document retrieved. We evaluate Bullseye with regard to three aspects: {\em system effectiveness}, {\em user effectiveness}, and {\em user effort}. In a system-blind evaluation, users were asked to compare passage retrieval using Bullseye vs.\ a baseline which ignores document structure, in regard to four types of graded assessments. Results show modest improvement in system effectiveness while both user effectiveness and user effort show substantial improvement. Users also report very strong demand for passage highlighting in scholarly search across both systems considered.

\end{abstract}

\category{H.3.4}{Information Systems}{Information Storage and
Retrieval, Systems and Software}

\terms{Algorithms, Experimentation}

\keywords{Focused Retrieval; Interactive Information Retrieval}

\section{Introduction}
\label{sec:Introduction}
\label{sec:relatedWorks}
While scholarly search engines like Google Scholar and Microsoft Academic Search enable quick and easy access to past research papers, search continues to be performed at the document level. This requires users to search through documents manually to find relevant material, or to determine that a document was not relevant after all. In contrast, {\em passage retrieval}~\cite{salton1993approaches,kaszkiel1997passage} and {\em focused retrieval}~\cite{trotman2007report} methods seek to identify relevant portions of the document to ease this burden. While one could return only passages, Kamps et al.~\cite{kamps2003xml} found out that users and assessors still regard whole articles as the meaningful unit of retrieval. Consequently, we develop and evaluate an approach of highlighting of relevant passages~\cite{knaus1996highlighting,milic2005facility,allan2005will} in scholarly documents.

In scholarly search, section structures in research papers provide important context for locating relevant information, which either the system may exploit in search, or the user may specify in expressing section-level relevance criteria for search. For example, Lalmas et al.~\cite{lalmas2007evaluating} distinguish {\em Content-only} (CO) INEX topics, in which users do not express structural relevance criteria, vs. {\em Content-and-structure} (CAS) topics, in which users express structural constraints. 
Our review of a few decades of past research papers suggests that common section headers have only changed minimally over time, suggesting an abundance of training data and generality of techniques. Extracting this implicit structure across varying document formats is an open challenge (cf.~\cite{denny2009evaluation}).

In this work, we investigate the following research questions in the specific domain of scholarly search. 
\begin{itemize}
\item How can we classify and localize implicit structure in scholarly documents for passage retrieval?  
\item How can we improve user satisfaction by combining content and structural features in passage-level retrieval?  
\item How can passage highlights improve user satisfaction?
\end{itemize}


The {\em Bullseye} system developed in this work adopts an off-the-shelf passage retrieval algorithm by MITRE~\cite{light2001analyses}, but uses detected section structure information to restrict results to sections specified by the user. To highlight relevant passages in PDF documents, we utilize the Apache PDFBox~\cite{PDFBox} library. While we had expected this to be straightforward, finding exact locations of the text in the PDF turned out to be non-trivial, as further discussed later.

Following Maskari et al.~\cite{al2010review}, we measure three aspects contribution toward user satisfaction: {\em system effectiveness}, {\em user effectiveness}, and {\em user effort}. In a system-blind evaluation, users were asked to compare Bullseye highlighting vs.\ structure-agnostic MITRE highlighting. System effectiveness was measured by adopting a focused retrieval metric $quant_{gen}$ from INEX~\cite{lalmas2007evaluating}, which combines two other measurements: \emph{specificity} and \emph{exhaustivity}. Users reported each on a graded scale. User effectiveness was measured by asking participants to report overall satisfaction on a graded-scale. Finally, user effort asked users to quantify the amount of time saved (or lost) on a graded scale. While results show only modest improvement in system effectiveness, both user effectiveness and user effort show substantial improvement.

\begin{figure}[!hbt]
\begin{center}
\ifpdf
\includegraphics[width=1.1\columnwidth, trim=8cm 0cm 0cm 5cm]{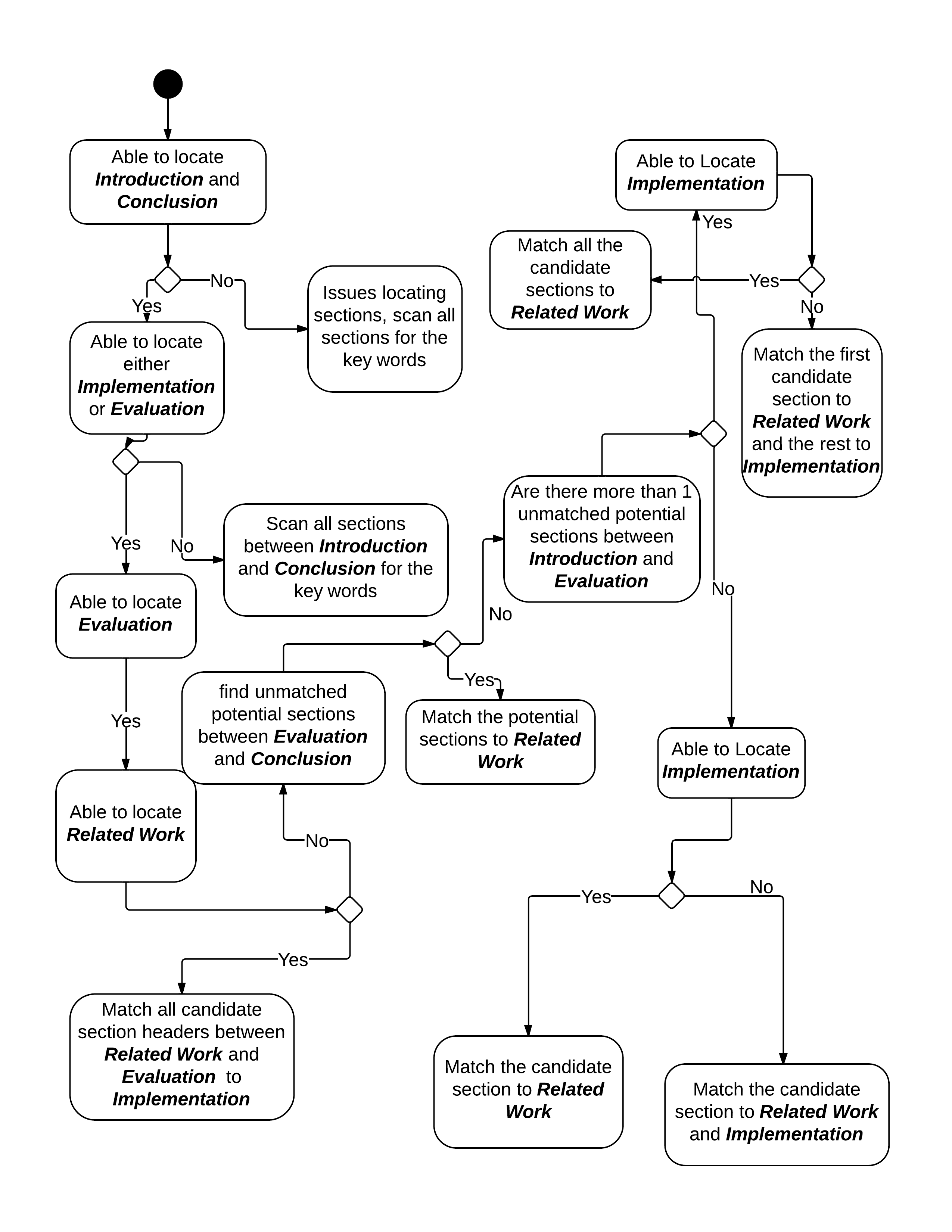}
\else
\includegraphics[width=1.1\columnwidth]{DecisionTree.png}
\fi
\end{center}
\vspace{-1.25cm}
\caption{Decision Tree}\label{fig:TE}
\vspace{-0.5cm}
\end{figure}

\section{The Bullseye System}


Given a collection of PDF documents, Bullseye first pre-processes documents to extract text, detect document sections, and map observed section headers to a fixed set of target sections. To search, Bullseye takes two key inputs: a search query and a listing of any document sections to which results should be restricted. Bullseye does not define a new passage retrieval algorithm but can incorporate any such method off-the-shelf. In this work, we adopt a very simple algorithm developed by MITRE~\cite{light2001analyses}, which treats each sentence as a passage. Following sentence boundary detection, stopwords are removed~\cite{frank2000text}, and terms are then stemmed~\cite{croft1994corpus} and matched between each query and its passage candidates. We then filter the set of retrieved passages so that any found outside the user's listing of specified sections are discarded. 

Finally, relevant passages in the PDF documents are highlighted using Apache PDFBox~\cite{PDFBox}. While we had expected this to be straightforward, finding exact locations of the text in the PDF turned out to be non-trivial. To accomplish this, we: 1) mark the beginning of each article, page, paragraph, and word; 2) determine the default text orientation in the document (e.g., the glyph is written from left to right or from right to left); determine the default document structure (e.g., single column vs multi column); determine the default font size and spacing (e.g., line spacing and word spacing); and finally 4) detect overlapping text.

\subsection{Detecting Document Structure}

Our approach is inspired by Denny et al.~\cite{denny2009evaluation}, whose {\em SecTag} algorithm identifies both labeled and un-labeled (implied) section document headers using Bayesian statistics. We first created a training set of 200 research papers from a variety of software engineering conferences: FSE, ICSE, ASE, ICCPS, MobiCom, and OOPSLA. Manual inspection showed that section headers can be fairly broadly defined by a fixed set of {\em Abstract, Introduction, Related Work, Implementation, Evaluation, Conclusion} and {\em Future Work} sections. 
Many papers we inspected had section headers which were either exact matches or equivalent terms (e.g. \emph{Summary} is equivalent to \emph{Conclusion/Future Work}). Some papers did have unusual section headers which do not match explicitly with headers in the target set and could not be clearly mapped. 
However, since users would likely not know to query for these sections either, this seems to be a relatively minor concern. 
The following observed patterns were incorporated into Bullseye (as deterministic decision trees) for matching observed section 
headers to target headers.

\begin{enumerate} [(1)]
 \item Section headers largely respect implicit ordering. 
 \item Observed section headers often exhibit a many-to-one relationship to our target set of sections. 
 \item Observed section headers tend to fall into one of three target section headers : \emph{Related Work}, \emph{Implementation} and \emph{Evaluation}.
 \item Not all target section headers are available in all papers. 
 \item \emph{Implementation} and \emph{Evaluation} target section often match multiple observed sections
 \item \emph{Related Work} is less likely to have multiple observed section headers matched to it, compared with \emph{Implementation} and \emph{Evaluation}.
\end{enumerate}

Bullseye first processes the document, sentence by sentence, to find all possible section headers. A section header is one that marks the beginning of sentences and consists of only Capital letters and a leading number or Roman numbers. 
All the possible section headers are inserted into a linked list which preserves the insertion order. Each section header candidate is matched against the equivalent terms for sections defined in the target set. The matched results are recorded into the linked list. Bullseye then checks in the linked list whether it can find matches for all the section search conditions specified by the user. If it could not locate all matching sections, Bullseye uses the decision trees based on the ordering and statistical information found previously to locate section headers specified by the user. 

Fig.~\ref{fig:TE} shows {\em one} of the learned decision trees in the Bullseye algorithm. This decision tree specifically is used to determine implicit \emph{Related Work} and \emph{Implementation} when \emph{Evaluation} section is found through explicit matching process.  The decision process
conforms to training set statistics. Through the explicit and implicit section header matching, Bullseye maintains a state machine of how many potential section headers have been explicitly matched and how many are implicitly matched. The intersection of matched section headers in the state machine and user specified section query condition are used to pinpoint those sections for locating the keywords submitted in the user query.

Note that we have built logic into Bullseye to detect massive section location matching failure and resort back to ignoring in this exceptional condition.

\section{Experimental Setup}
\label{sec:evaluationSettings}

Following Maskari et al.~\cite{al2010review}, we measured three aspects contribution toward user satisfaction: system effectiveness, user effectiveness, and user effort. In a blind evaluation, users were asked to use and evaluate each system, our Bullseye system and the Baseline system. We also asked users how beneficial passage-level highlighting is for scholarly search.

{\bf System effectiveness} was measured by adopting a focused retrieval metric $quant_{gen}$ from INEX~\cite{lalmas2007evaluating}, which combines two other measurements: \emph{specificity} and \emph{exhaustivity}. Specificity measures how focused the retrieved information is on relevant material and free from other extraneous material. For each highlighted passage retrieved by the system, participants evaluated it on a 4-point graded scale:
\begin{enumerate} [1:]
\setcounter{enumi}{-1}
\item Not specific: the topic of request is not a theme discussed in the element.
\item Marginally specific: the topic of request is a minor theme discussed in the element.
\item Fairly specific: the topic of request is a major theme discussed in the element.
\item Highly specific: the topic of request is the only theme discussed in the element.
\end {enumerate}
Exhaustivity, on the other hand, measures how fully the retrieved material covers all aspects of the underlying information need.  Participants judged it on a 3-point scale:
\begin{enumerate} [1:]
\setcounter{enumi}{-1}
\item Not exhaustive: the element did not discuss the query.
\item Partly exhaustive: the element discussed only few aspects of the query.
\item Highly exhaustive: the element discussed most or all aspects of the query.
\end{enumerate} 
$Quant_{gen}$ slightly favors exhaustivity, assigning high scores to exhaustive, but not necessarily specific elements: 
\[
 quant_{gen} =
  \begin{cases}
   1 	    &  if (e, s) = (3, 3) \\
   0.75   &  if (e, s) \in \{(2, 3), (3, \{2, 1\})\};\\
   0.5     &  if (e, s) \in \{(1, 3), (2, \{2, 1\})\};\\
   0.25   &  if (e, s) \in \{(1, 2), (1, 1)\};\\
   0 	    &  if (e, s) = (0, 0) \\
  \end{cases}
\]

{\bf User effectiveness} was measured by asking participants to report overall satisfaction with the given system being used for ascertaining the relevance of a given document. Users were asked to evaluate satisfaction on a 3-point scale: {\em not satisfied} (1), {\em slightly satisfied} (2), or {\em highly satisfied} (3). 

{\bf User effort} quantifies effort needed to find relevant results (e.g., number of clicks, the number of queries and the number of query reformulations, or rank position accessed to obtain relevant information. We asked participants to quantify their effort in terms of search time required to find the relevant information sought. Participants were asked to report their effort on a 4-point graded scale: {\em highly reduced} (1), {\em slightly reduced} (2), {\em no change} (3), or {\em increased} (4).  

{\bf Highlighting preference.} We also asked users whether integrating such highlighting functionality with Google Scholar search results would improve their satisfaction?  Users were asked this same question regardless of system used in the blind evaluation, using a 4-point scale: {\em worse} (1), {\em no difference} (2), or {\em slightly better} (3), or (4) {\em much better.}

\subsection{User Study}

The participant pool was comprised of 15 university students from Software Engineering, Computer Science and Information Science disciplines. Each provided a personal test collection (document sets, content and structural queries, and manually judged set of gold passages). We restricted documents submitted by participants to be within 6-12 pages. Because users familiar with search topics might judge system effectiveness more prudently than unfamiliar users, this could potentially introduce a confounding variable into the evaluation. To mitigate this risk, each also evaluated 5 other query and document sets provided by a different user.

For each participant's submission (document and query), we ran the baseline algorithm and Bullseye to create two separate documents (Algorithm Documents). For each topic (all together 20 topics), there was a set (Test Set) which contains the original document, query, ``gold'' true passages and these two Algorithm Documents generated by the algorithms. 
Each participant was given six Test Sets, with one set containing his own submission. Users then evaluated each Algorithm Document based on the ``gold'' true passages, the document and the query. 

Before the user study, we organized a training session with all participants reviewing aims of our study, evaluation metrics, evaluation process, and participant responsibilities. 
Each participant was made aware that the Test Set to be provided consisted of two Algorithm Documents with relevant passages highlighted by two alternative algorithms. We explained evaluation questions being asked. In this manner, each user was trained how to create gold set by himself and compare the gold set with the algorithms' output.
 

\section{Results}
\label{sec:evaluation}

\begin{figure}[htb]
\captionsetup[subfigure]{aboveskip=-1pt,belowskip=-1pt}
\centering
\begin{subfigure}[b]{2.7cm}{
 \includegraphics[width=2.7cm]{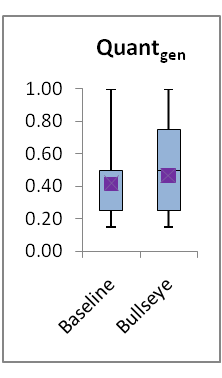}
 \caption{$quant_{gen}$}
 \label{fig:Quant}}
\end{subfigure}
\begin{subfigure}[b]{2.7cm}{
  \includegraphics[width=2.7cm]{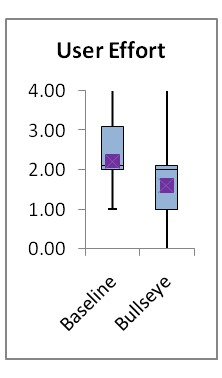}
   \caption{Effort}
  \label{fig:UserEffort}}
\end{subfigure}
\begin{subfigure}[b]{2.7cm}{
  \includegraphics[width=2.7cm]{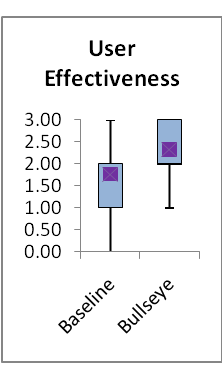}
  \caption{Effectiveness}
  \label{fig:UserEffectiveness}}
\end{subfigure}
\caption{Evaluation Metrics}
\label{fig:figure}
\vspace{-0.7cm}
\end{figure}

{\bf System Effectiveness.} Fig.~\ref{fig:Quant} shows the box-and-whisker plot of $quant_{gen}$ score distributions for Bullseye vs.\ the Baseline (no structural filtering). While Bullseye's average improvement is modest (0.47 vs.\ 0.41 mean), its score distribution skews much higher. The lower quartile (Q1) is almost equivalent for the Bullseye and Baseline (0.26 vs.\ 0.25). but in the upper quarter (Q3), Bullseye is much better (0.75 vs.\ 0.5). 
In some cases, though, sentences matching the query term happen to exist only in the user-specified sections, in which case structural filtering has no effect. 

{\bf User Effort.} Fig.~\ref{fig:UserEffort} shows the box-and-whisker plot score distribution of Bullseye vs.\ Baseline. We see that Bullseye substantially reduces user effort vs.\ the Baseline, indicating users reported far less effort involved when using Bullseye in the blind evaluation. Reductions include the mean (1.61 vs.\ 2.22), Q1 (1.0 vs.\ 2.0), and Q3 (2.0 vs.\ 3.0). 

{\bf User Effectiveness.} Fig.~\ref{fig:UserEffectiveness} shows the box-and-whisker plot score distribution for Bullseye vs.\ the Baseline. We see that users report being much more satisfied with Bullseye in the blind evaluation in the mean (2.33 vs.\ 1.76), Q1 (2.0 vs.\ 1.0), and in Q3 (3.0 vs.\ 2.0). 

{\bf Highlighting preference.} Participants were unanimous across both systems in indicating that integrating such highlighting would yield much better satisfaction. While we do not learn anything about relative performance of one system vs.\ the other from this (and so do not show a corresponding boxplot), we observe clear and very strong demand for supporting passage-highlighting in scholarly search. 

%
%

\section{Limitations}
\label{sec:validity}

We have assumed a fix set of target sections to match, training our system entirely on examples from the Software Engineering domain. We have assumed test documents will be similar to training documents. While our participants evaluated our system on papers from wider Information and Computer Science domains, our current system remains very far from being able to support all scholarly genres and document formatting and structure. Our evaluation shows in the case of mismatch between training and testing documents used here, Bullseye appears to exhibit at least comparable performance to simply ignoring document structure, as done in the baseline system, given our fallback design.

Another issue is that the decision trees used by the Bullseye reflect our own empirical inspection of the training set. While we also applied two automatic decision tree induction techniques~\cite{umanol1994fuzzy}~\cite{ruggieri2002efficient}, the trees we constructed ourselves proved more effective in practice. That said, future work should develop an effective automated method to reduce manual effort and strive for greater robustness and generality over a wider range of scholarly domains. 




\section{Conclusion and Future Work}
\label{sec:FutureWork}
We are motivated by the fact that a user satisfaction can be enhanced by improving the overall search experience. We delivered this by implementing Bullseye algorithm which is able to 
narrow down the search within user-specified sections and highlight highly relevant passages within the document. The results show Bullseye is able to improve
user satisfaction and search effectiveness. 

We wish to port Bullseye algorithm to a Google Chrome Application so after a user retrieves the documents from a search engine (e.g. Google Scholar), the browser automatically highlight the most relevant information to the user. 



\section*{Acknowledgments}

This study was supported in part by National Science Foundation grant No. 1253413, DARPA Award N66001-12-1-4256, and IMLS grant RE-04-13-0042-13. Any opinions, findings, and conclusions or recommendations expressed by the authors are entirely their own and do not represent those of the sponsoring agencies


\bibliographystyle{plain}
\bibliography{library}

\end{document}